\documentclass[12pt,preprint]{aastex}

\received{2001 August 6, revised version 7 September 2001 / 20 October 2001}
\begin{document}

\title{NEUTRINOS FROM EARLY-PHASE, PULSAR-DRIVEN SUPERNOVAE} 
\vspace{1truecm}
\author{J.H. Beall \altaffilmark{1,2,3} \& W. Bednarek \altaffilmark{4}}

\altaffiltext{1}{E.O. Hulburt Center for Space Research, Naval Research
Laboratory, Washington, DC 20375}
\altaffiltext{2}{Center for Earth Observing and Space Research, School 
for Computational Sciences, George Mason University, Fairfax, VA 22030}
\altaffiltext{3}{St. John's College, Annapolis, MD 21404, USA}
\altaffiltext{4}{Department of Experimental Physics, University of \L
\'od\'z, ul. Pomorska 149/153, PL 90-236 \L \'od\'z, Poland }

\begin{abstract}

Neutron stars, just after their formation, are surrounded by expanding,
dense, and very hot envelopes which radiate thermal photons. Iron nuclei can
be accelerated in the wind zones of such energetic pulsars to very high
energies. These nuclei photo-disintegrate and their products lose energy
efficiently in collisions with thermal photons and with the matter of the
envelope, mainly via pion production. When the temperature of the radiation
inside the envelope of the supernova drops below $\sim 3\times 10^6$ K,
these pions decay before losing energy and produce high energy neutrinos. 
We estimate the flux of muon neutrinos emitted during such an early phase of
the pulsar - supernova envelope interaction. We find that a 1 km$^2$
neutrino detector should be able to detect neutrinos above 1 TeV within
about one year after the explosion from a supernova in our Galaxy.  This
result holds if these pulsars are able to efficiently accelerate nuclei to
energies $\sim 10^{20}$ eV, as postulated recently by some authors for
models of Galactic acceleration of the extremely high energy cosmic rays
(EHE CRs).

\end{abstract}

\noindent
%{\it Subject headings:}


%Submitted to the Astrophysical Journal Letters - August 6th, 2001
%Revision submitted - October 21st, 2001

\eject
\section{Introduction}
The production of neutrinos with different energies during supernova
explosions has been discussed extensively during the last several years,
mainly in the context of gamma-ray bursts (GRBs). For example, neutrinos
with energies $> 100$ TeV can be produced in the interactions of protons
accelerated by a fireball shock with GRB photons (e.g. Waxman \&
Bahcall~1997, Vietri~1998). TeV neutrinos can arise in interactions of
protons with radiation when the fireball jet breaks through the stellar
envelope (Meszaros \& Waxman~2001). GeV neutrinos can also be produced in
the interactions of protons with neutrons which can be present in the
fireball models discussed by Derishev et al.~(1999) and Meszaros \&
Rees~(2000). 

Of course, the acceleration of particles to high energies is also expected
in the case of classical supernovae.  Recently, Waxman \& Loeb~(2001)
estimated the neutrino flux from a Type II supernova when the shock breaks
out of its progenitor star.  Berezinsky \& Prilutsky~(1978) and 
Protheroe, Bednarek \& Luo~(1998) have estimated the flux of neutrinos
produced by particles accelerated by the young pulsar during the early phase
of the  supernova explosion (see also the calculations for the Crab Nebula
case in Bednarek \& Protheroe~1997).

In this paper we show that neutrinos can also be produced soon after the
pulsar formation inside the supernova envelope and that these neutrinos are
detectable by current neutrino detectors.  As shown by Blasi, Epstein \&
Olinto~(2000, BEO) or De Goubeia Dal Pino \& Lazarian~(2000), among others,
particles can be accelerated to very high energies within the
``plerion-like" region of a supernova shortly after the formation of the
neutron star.  We consider the scenario in which iron nuclei accelerated above
the light cylinder of the pulsar interact:
first, with the thermal radiation of the expanding supernova envelope, and
later, with the matter of the envelope.  Acceleration of hadrons in the
pulsar wind zone has been discussed in the context of the production of high
energy cosmic rays since shortly after the discovery of pulsars (e.g. Gunn
\& Ostriker~1969, Karaku\l a, Osborne \& Wdowczyk~1974).  For likely
parameters of pulsars at birth, we predict that the flux of muon neutrinos
can be observable by large-sized neutrino detectors during about one
year after supernova explosion.

\section{The physical scenario}

We consider type Ib/c supernovae, whose progenitors are Wolf-Rayet
type stars. Such stars evolve from massive stars with $M \ge 35$
M$_{\odot}$, and create iron cores surrounded by relatively light envelopes
of the order of a few solar masses. We use the models for the evolution
of such stars and their explosions as published by Woosley et al.~(1993). As
an example, we concentrate on their model 60 WRA.  The iron core collapses
to a very hot proto-neutron star which cools to the neutron star during about
$t_{\rm NS}\approx 5-10$s from the collapse (Burrows \& Lattimer~1986,
Wheeler et al.~2000). The rest of the mass of the presupernova (the
envelope) is expelled with the velocity at the inner radius of the order of
$v_1 = 3\times 10^8$ cm s$^{-1}$. However because of the density gradient,
the outer parts of the envelope move faster. We approximate the velocities
of matter in the envelope by the profile
\begin{eqnarray} 
v(R) = v_1 (R/R_1)^b, 
\label{eq1}
\end{eqnarray}
\noindent
where the parameter $b = 0.5$ is obtained from an approximation of the
velocity profile in the expanding envelope shown in Fig.~9 in Woosley et
al.~(1993), and we use $R_1 = 3\times 10^8$ cm as the inner radius of the
envelope at the moment of explosion. The density of matter in the envelope
just before the collapse of the iron core can be approximated by the profile
\begin{eqnarray}
 n(R) = n_1 (R/R_1)^{-a}, 
\label{eq2}
\end{eqnarray}
\noindent
where the density at $R_1$ is  $n_1 = 1.2\times
10^{31}$ cm$^{-3}$, and the parameter $a = 2.4$ is obtained
from Fig.~1 in Khokhlov et al.~(1999) by interpolation of the profiles for
the radius and density versus the mass of the presupernova star (see also
Fig.~3c in Woosley et al.~1993). The initial column density decreases with
time, t, due to the expansion of the envelope according to 
\begin{eqnarray}
\rho (t) = \int_{R_1}^{R_2} n(R) \left({{R}\over{R + v(R)t}}\right)^2 dR,
\label{eq3}
\end{eqnarray}
\noindent
where $R_2 = 3\times 10^{10}$ cm is the outer radius of the envelope at the
moment of explosion, and $v(R)$ and $n(R)$ are given by Eqns.~\ref{eq1} and
\ref{eq2}, respectively. Just after the collapse of the iron core the
temperature at the bottom of the envelope is $T_{0}\approx 3\times
10^9$ K at $R_1$ (see Fig.~8 in Woosley et al.~1993). Therefore the volume
above the pulsar and below the expanding envelope is filled with thermal
radiation which is not able to escape because of the high optical depth of
the envelope. We apply that the temperature of this radiation drops with
time during the expansion of the envelope according to
\begin{eqnarray}
T(t) = T_{0} \left({{R_1}\over{R_1 + v_1(t_{\rm NS} + t)}}\right)^{3/4}.
\label{eq4}
\end{eqnarray}
\noindent
At the moment of the neutron star's formation (after $\sim 10$ s), the
temperature in the region below the envelope already drops to $\sim 5\times
10^8$ K.

We assume that at this early age, the pulsar loses energy only via
electromagnetic radiation. Therefore, its period changes according to the
formula
$P_{\rm ms}^2(t) = 1.04\times 10^{-9} tB_{12}^2 + P_{0,ms}^2$, where
$P_{\rm 0,ms}$ $P_{\rm ms}$ are the initial and present periods of 
the pulsar (in milliseconds),
and $B_{12}$ is the pulsar surface magnetic field in units $10^{12}$ G. 
Note that after about 1 yr, the pulsar may also lose
efficiently energy on emission of gravitational radiation due to the r-mode
instabilities which are excited in cooling neutron star (e.g.
Andersson~1998, Lindblom, Owen \& Morsink~1998). During this time, it is
likely that the pulsar period changes suddenly reaching the value $\sim 10 -
15$ ms at about 1 yr after formation.

During the first year after explosion the rate of rotational energy loss by
the pulsar is too low to influence the initial expansion velocity of the
envelope (Ostriker \& Gunn~1971), assuming that the pulsar has been born
with parameters characteristic for the classical radio pulsars
(e.g. the Crab pulsar). Only pulsars with periods of the order of a
few milliseconds and super-strong magnetic fields (magnetars) can significantly
accelerate the envelope at short time intervals after the explosion. 

In the next section, we consider the acceleration of the iron nuclei in the
pulsar magnetosphere above the light cylinder and below the expending envelope,
adopting the above model for the pulsar formation and expansion of the
supernova envelope.  Our aim is to find out if the acceleration and
radiation processes inside the expanding envelope can produce an observable
flux of neutrinos during the early phase of supernova explosion.

\section{Acceleration of iron nuclei}

\begin{figure}[b] 
  \vspace{5.5cm} 
 \includegraphics{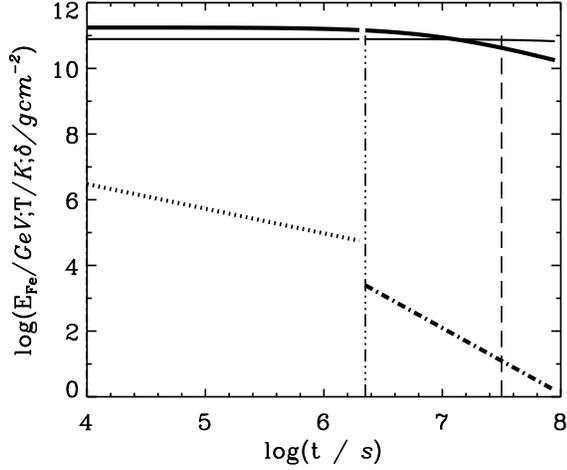}
\caption[]{The dependences of the maximum energies of accelerated iron nuclei,
the temperature of thermal radiation in the acceleration region
(dotted curve), and the column density of
matter in the envelope (thick dot-dashed curve) are shown as a function
of time which is measured from the supernova explosion. 
Iron nuclei are accelerated
by the pulsars with initial parameters: $B_{12} = 4$, $P_{\rm ms} = 3$
(thin full curve) and $B_{12} = 100$, $P_{\rm ms} = 10$ (thick full 
curve). The vertical lines mark the times:
after which the envelope becomes transparent to thermal radiation
(dot-dot-dot-dashed), and at which the envelope becomes transparent to
hadronic collisions and the parameters of the neutron star may change
drastically due to the gravitational energy losses (long dashed).}         
\label{fig:prof}  
\end{figure}

Following the recent work by Blasi, Epstein \& Olinto~(BEO), we assume that
the magnetic energy in the pulsar's wind zone accelerates iron nuclei close
to the light cylinder radius in the mechanism called magnetic slingshot
(Gunn \& Ostriker~1969). Since this acceleration occurs very fast, the
nuclei photo-disintegrate and lose energy during farther propagation in the
radiation field below the supernova envelope (the cooling phase) but not
during acceleration process. The energy that the iron nuclei can reach in
the wind zone does depend on the pulsar parameters, so that

\begin{eqnarray}
E_{Fe} = {{B^2(r_{\rm LC})}\over{8\pi n_{\rm GJ}(r_{\rm LC})}}  
\approx 1.8\times 10^{11} B_{12} P_{ms}^{-2}~{\rm GeV}, 
\label{eq5}
\end{eqnarray}

\noindent
where $r_{\rm LC} = cP/2\pi$ is the light cylinder radius, and $n_{\rm GJ} =
B(r_{\rm LC})/(2 e Z c P)$ is the Goldreich \& Julian (1969) density at the
light cylinder.  As an example we show the temperature of the radiation
field below the envelope of supernova and the energies of accelerated iron
nuclei as a function of time in Fig.~\ref{fig:prof}, in the case of two
pulsars with parameters: $B_{12}=4$, $P_{\rm ms}=3$, and $B_{12}=100$,
$P_{\rm ms}=10$.  Note that in the case of the first pulsar, its period does
not change, and consequently, the energies of the iron nuclei do not change
significantly during the first year after the explosion.  This is not the
case for the second pulsar.

\clearpage
\clearpage

We obtain the spectrum of iron nuclei accelerated in the pulsar wind zone close
to its light cylinder by following general prescription given by Blasi,
Epstein \& Olinto~(BEO). In this simple model all nuclei are accelerated to
the energy, $E_{\rm Fe}$ and their number is a part of the Goldreich \& Julian 
density, $n = \xi n_{\rm GJ}$. However, contrary to that
work, we assume that the pulsar with specific parameters ($B_{12}$ and
$P_{\rm ms}$) injects particles within some range of energies, due to the
fact that the magnetic field at different parts of the light cylinder radius
(and so the Poynting flux)
is different. The magnetic field strength at the height $h$, measured 
from the plane containing the pulsar and perpendicular to the light
cylinder radius, can be expressed by 
$B(r)\approx B(r_{\rm LC})\cos\alpha$,
where $\cos\alpha = r_{\rm LC}/(r_{\rm LC}^2 + h^2)^{1/2}$.
Therefore, the density of particles at
the light cylinder $n(h) = n_{\rm GJ}\cos^3\alpha$, and their energies, $E =
E_{\rm Fe}\cos^3\alpha$ depend on $h$. We calculate the number of
particles injected at the height $h$ per unit time from $dN/dt = 2\pi r_{\rm
LC} c n(h) dh$. Using the above formulae, we replace $dh$ by $dE$ and obtain
the  differential spectrum of iron nuclei injected by the pulsar at the
fixed age of the pulsar, t,
\begin{eqnarray}
{{dN}\over{dEdt}} = {{2\pi c \xi r_{\rm LC}^2 n_{\rm GJ}(r_{\rm LC})
(E_{\rm Fe} E^2)^{-1/3}}\over{3\left[(E_{\rm Fe}/E)^{2/3} -
1\right]^{1/2}}}  \cong  {{3\times 10^{30}\eta (B_{12} P_{\rm
ms}^{-2}E^{-1})^{2/3}} \over{\left[(E_{\rm Fe}/E)^{2/3} - 1\right]^{1/2}}}
~{\rm {{Fe}\over{s~GeV}}}.     
\label{eq9}
\end{eqnarray} 
\noindent
Note that our parameter $\xi$ 
has similar meaning to the parameter $\xi$ introduced by Blasi et al. (BEO).
Due to the shape of the spectrum of iron nuclei $\propto E^{-1/3}$
(Eq.~\ref{eq9}), also in our model most of the iron nuclei reaches the energies
from the highest energy part of this spectrum at $E_{\rm Fe}$, given by 
Eq.~\ref{eq5}. 

In this paper we discuss only the consequences of acceleration of iron
nuclei at a relatively early phase after the supernova explosion, i.e up to
about one year from pulsar formation. During this time, the radiation field inside
the expanding supernova envelope, and thereafter, the column density of the
envelope, are high enough to provide a target for relativistic nuclei. As
we have already noted, the period of the neutron star can be significantly
influenced by gravitational energy losses after about 1 year from the time
of the explosion, when the neutron star cools enough.

\section{Production of neutrinos}

Accelerated nuclei move in the pulsar wind almost at rest in the wind
reference frame (BEO). Therefore they do not lose
significant energy by synchrotron emission.  As we have noted, however,
these nuclei will interact with the strong thermal radiation field in the
supernova cavity, suffering multiple photo-disintegration of nucleons.
Since the mean free paths for photo-disintegration of iron (and lighter) nuclei
are significantly shorter than for their energy losses on $e^\pm$ pair and pion
production (see e.g. Karaku\l a \& Tkaczyk~1993), the nuclei suffer complete
disintegration onto nucleons before significant energy losses.
For plausible parameters of the pulsar and the acceleration region, these
secondary nucleons lose energy mainly via pion production. The
pions then decay into high energy neutrinos if their decay distance scale
$\lambda_\pi \approx 780\gamma_\pi~~{\rm cm}$, 
is shorter than their characteristic energy loss mean free path. The Lorentz
factors of pions, $\gamma_\pi$, are comparable to the Lorentz factors of
their parent protons, so they move similarly in the pulsar wind and their
synchrotron losses should not dominate over their inverse Compton losses
(ICS) in the thermal radiation. Pions lose energy on ICS process mainly in
the Klein-Nishina (KN) regime, but not very far from the border with the
Thomson regime. Therefore we can estimate the ICS
losses of pions in the KN regime by
\begin{eqnarray}
P^{\rm ICS}_{\rm KN}\approx 4/3 \pi \sigma_{T} U_{\rm rad} (m_e/m_\pi)^2 
\gamma_{\rm KN/T}^2, 
\label{eq10}
\end{eqnarray}
\noindent
where $\sigma_{\rm T}$ is the Thomson cross section, $U_{\rm rad}$ is the
energy density of radiation, $m_{\rm e}$, $m_\pi$ are the masses of electron
and pion, and $\gamma_{\rm KN/T}\approx 5\times 10^{11}/T$ is the Lorentz
factor at the transition between the KN and T regimes. We estimate the
mean free path for pion energy losses via ICS  
\begin{eqnarray}
\lambda_{\rm ICS}\approx m_\pi\gamma_\pi/P^{\rm ICS}_{\rm KN}\approx
10^{16}\gamma_\pi/T^2~{\rm cm}.
\label{eq11}
\end{eqnarray}
$\lambda_{\rm ICS}$ is comparable to the pion decay distance, $\lambda_\pi$,
only for temperatures of radiation $T \le 3\times 10^6$ K. So then pions
decay before significant energy losses only if this condition is fulfilled.

The temperature of the radiation inside the envelope drops to $T \le 3\times
10^6$ K at about $t_{\rm dec}\sim 10^4$ s after the supernova explosion (see
Eq.~\ref{eq4}). At that moment nucleons from desintegration of nuclei 
cool in collisions with thermal radiation mainly by pion production.
However, when the optical depth through the expanding envelope drops below
$\sim 10^3$, the radiation is not further confined in the region below the
envelope and its temperature drops rapidly.  Based on Eqs.~\ref{eq1},
\ref{eq2}, and~\ref{eq3}, we have found that this happens at the time $t_{\rm
conf}\sim 2\times 10^6$ s after the explosion (see the 
thin dot-dot-dot-dashed line in Fig.~\ref{fig:prof}). Therefore, we
conclude that nucleons are able to cool efficiently in the thermal radiation
and produce pions, which then decay into muon neutrinos, but only from
$t_{\rm dec}\approx 10^4$ s up to $t_{\rm conf}\approx 2\times 10^6$s after
the explosion. At later times, the relativistic iron nuclei do not
desintegrate in the radiation field but  interact directly with the
matter of the envelope whose density is already low enough so that pions
produced by that interaction are able to decay into neutrinos and muons.

\begin{figure}[t] 
  \vspace{10.cm} 
  \includegraphics{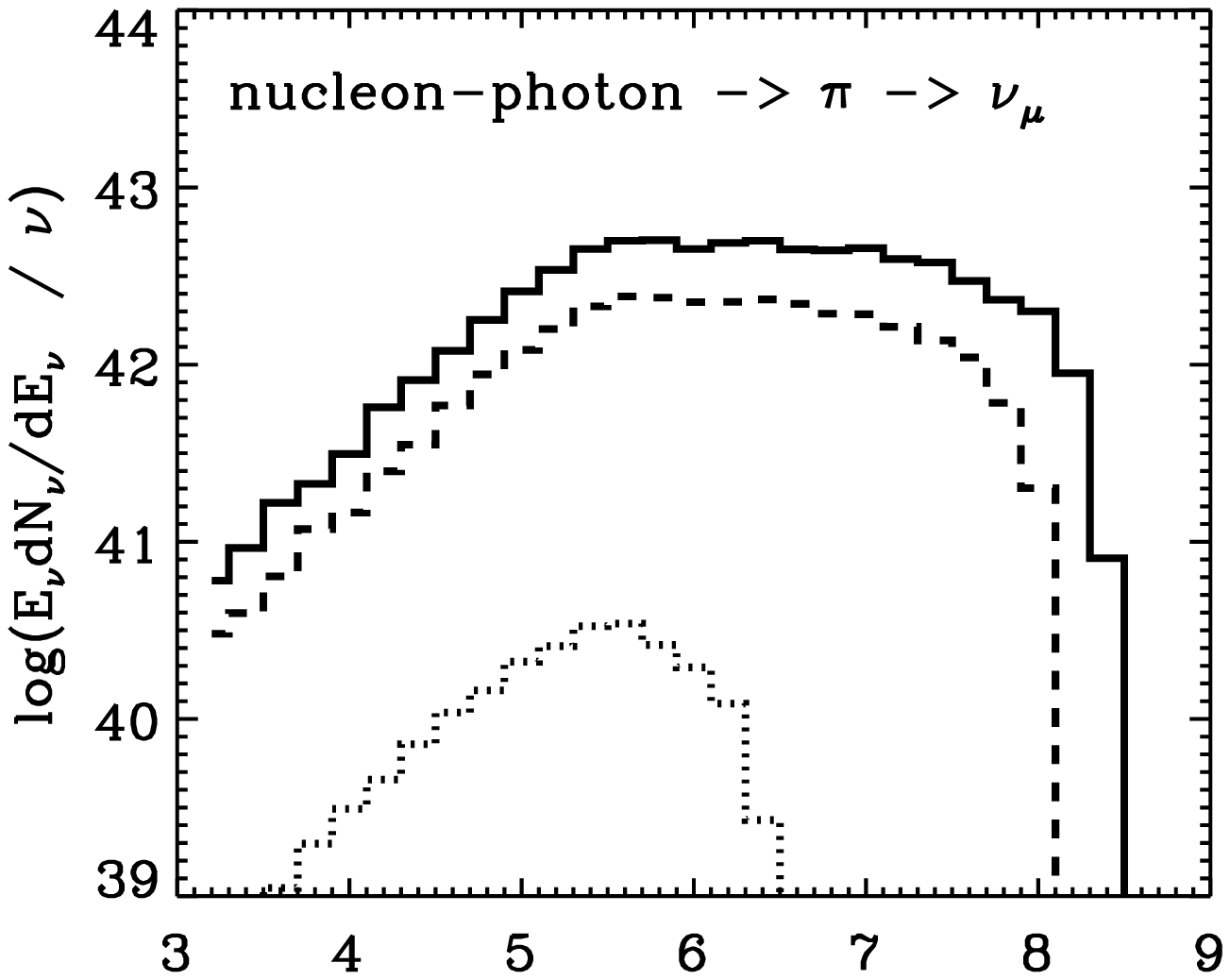}
  \includegraphics{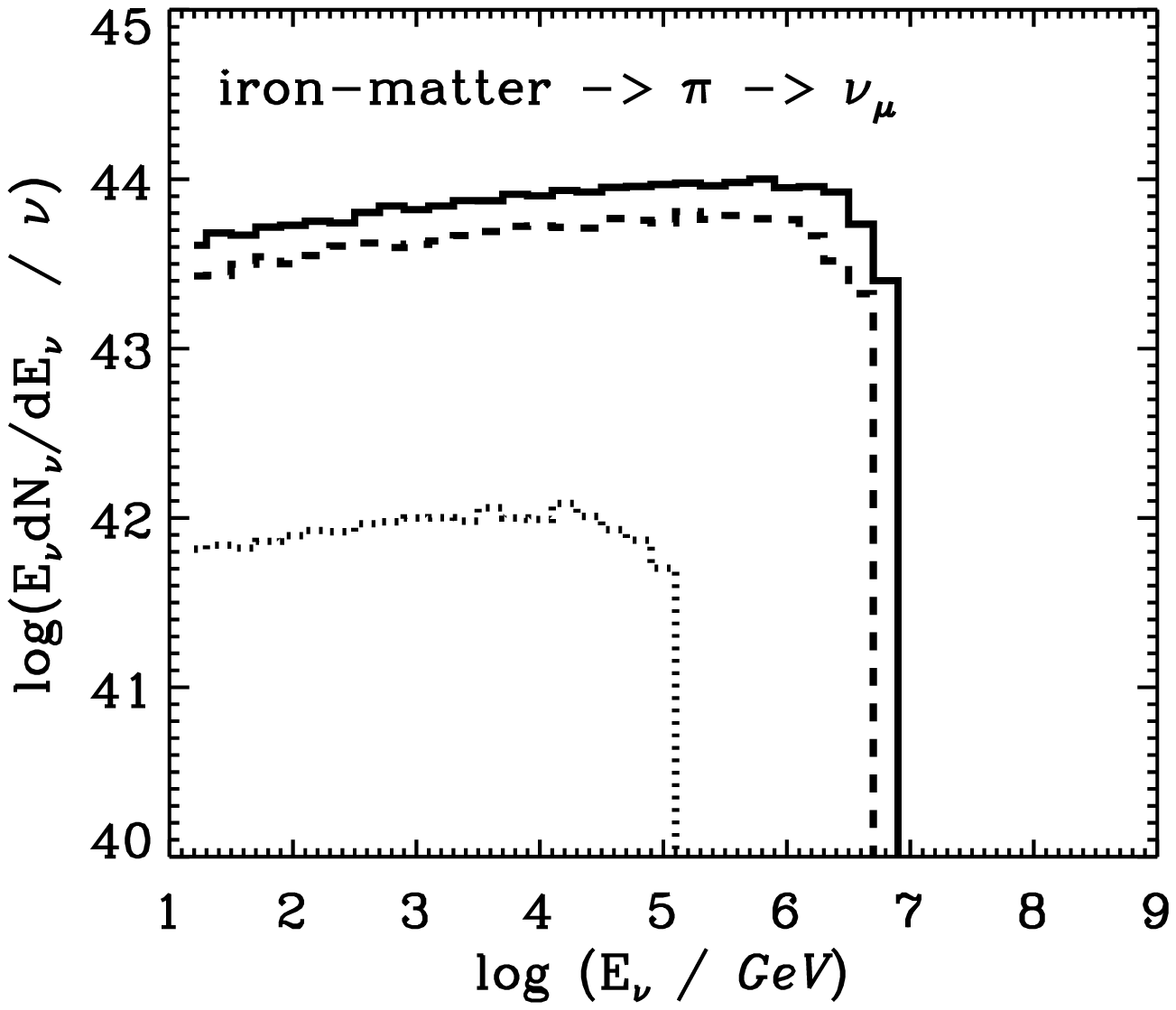}
\caption[]{Spectra of muon neutrinos and antineutrinos produced in interactions
of nucleons from photodesintegration of iron nuclei with the thermal
radiation field inside the supernova envelope
($N-\gamma\rightarrow\pi\rightarrow\nu_\mu$) and from interactions of iron
nuclei with the matter of the envelope
($Fe-M\rightarrow\pi\rightarrow\nu_\mu$). The density factor is $\xi = 1$,
and the initial periods and the surface magnetic fields of the pulsars are:
$P_{\rm ms} = 10$ and $B_{12} =100$ (full histograms), $P_{\rm ms} = 3$ and
$B_{12} = 4$ (dashed), and  $P_{\rm ms} = 20$ and $B_{12} = 4$ (dotted).}     
       \label{fig:neu-spect} \end{figure}

We now compute the differential spectra of muon neutrinos produced in the
interaction of nuclei: (1) with the radiation field below the envelope
during the period $1\times 10^4 - 2\times 10^6$ s after the supernova explosion; and
(2) with the matter of the envelope during the period from 
$2\times 10^6 - 3\times 10^7$s after the explosion, assuming that the
nucleons cool to the lowest energies allowed by the column densities of
photons and matter, respectively. In this calculation, we assume that pions
are produced in N-$\gamma$ collisions with the Lorentz factors comparable to
their parent nucleons. In the case of iron-matter (Fe-M) interactions, 
we apply the pion multiplicities given in Orth \& Buffington~(1976).  As an
example we show the results of these calculations (Fig.~\ref{fig:neu-spect})
for three pulsars with initial
parameters: $P_{\rm ms}=3$ and $B_{12}=4$ (model I, dashed histogram),
$P_{\rm ms}=20$ and  $B_{12}=4$ (model II, dotted histogram), and $P_{\rm
ms}=10$ and $B_{12}=100$ (model III, full histogram). The iron nuclei
accelerated by pulsars according to the models I and III fulfil the condition 
$P_{\rm ms}\approx 4 (B/10^{13} G)^{1/2}$ (given by BEO), 
which allow to reach energies $10^{20}$ eV. Note that even in the
case of the model III, the energy losses of the pulsar during first year after
its formation do not overcome the total kinetic energy of the envelope. 
Therefore the envelope is not accelerated.
Model II describes the pulsar with the presumable parameters of the Crab
pulsar at birth.  The periods of pulsars in models I and II do not change
drastically during the first year after the supernova explosion.  
However the period of the pulsar in model III changes by a factor of $\sim 3$
which causes the change in energy of the iron nuclei by about an order of
magnitude during one year after explosion.

The numbers of neutrinos produced in these two processes do differ
significantly, due to the fact that pion multiplicities in N-$\gamma$ and Fe-M
interactions are different.  Therefore the neutrino fluxes from Fe-M
interactions are up to an order of magnitude higher than the neutrino fluxes
from N-$\gamma$ interactions.  Note also that the column density of the
envelope after $\sim 2\times 10^6$s drops rapidly with time and only nuclei
injected earlier than $\sim 10^7$s can undergo multiple interactions with
matter.

\begin{figure}[b] 
  \vspace{5.5cm} 
 \includegraphics{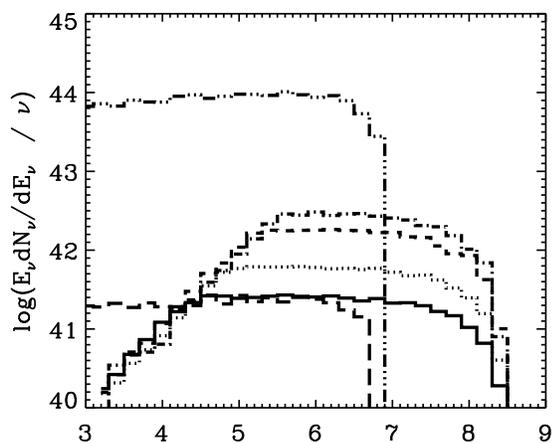}
\caption[]{Spectra of muon neutrinos and antineutrinos produced by the pulsar
with the parameters of the model III ($B_{12} = 100$, and $P_{\rm ms} = 10$) 
at different range of time measured from the pulsar formation: $\Delta t =
10^4 - 10^5$ s (full histogram), $10^5 - 3\times 10^5$ s (dotted), 
$3\times 10^5 - 10^6$ s (dashed), $10^6 - 2\times 10^6$ s (dot-dashed), 
$2\times 10^6 - 10^7$ s (dot-dot-dashed), and $10^7 - 3\times 10^7$ s 
(long dashed).}            
\label{fig:spec-time}  
\end{figure}

Since in model III the energies of accelerated iron nuclei significantly 
change during the first year it is interesting to investigate how the
spectrum of neutrinos changes with time in this case. In
Fig.~\ref{fig:spec-time} we show the spectra of neutrinos for the model III
produced at different range of time after pulsar formation.  
The neutrino spectra have comparable intensities before $\sim 2\times 10^6$ s
after explosion (during N-$\gamma$ production phase) with the cut-offs at this
same energy (the pulsar period do not change significantly during this time 
see Fig.~\ref{fig:prof}). However the location of the lower energy break
in the neutrino spectra 
shifts with time to higher energies since it is determined by the
temperature of radiation which drops with time. The highest fluxes of neutrinos
are expected at $\sim 2\times 10^6 - 10^7$ s after explosion during the
interaction of iron nuclei with the matter of the supernova envelope. 
At later times ($10^7 - 3\times 10^7$ s) the neutrino flux
drops significantly, because particles, 
accelerated already to lower energies, are not completely cooled in
collisions with the matter of the envelope (see Fig.~\ref{fig:prof}).   

Just after the collapse of the iron core, the column density of matter is
high enough to absorb the neutrinos with energies above $\sim 0.1$ GeV.
However, the column density drops quickly with time ($\propto t^{-2}$, see
Eq.~\ref{eq4}) and the cross-section for neutrino interaction with
matter increases with energy $\propto E_\nu$ and $\propto E_\nu^{0.4}$ below
and above $E_\nu\sim 1$ TeV, respectively (e.g. Hill~1997). Therefore the
optical depth becomes less than one at times $\sim 10^4$ s even for the
highest energy neutrinos produced.

\clearpage
\clearpage

\clearpage
\clearpage

\section{Discussion and conclusion}

For a supernova inside our Galaxy at a distance, D, $= 10$ kpc, we estimate
the expected flux of muon neutrinos produced (in nucleon-photon interactions
during $1\times 10^4 - 2\times 10^6$ s after the explosion and produced in
nuclei-matter collisions during $2\times 10^6 - 3\times 10^7$ s after the
explosion) by integrating the neutrino spectra shown in Fig.~2.  The
likelihood of detecting these neutrinos by a detector with a
surface area of 1 km$^2$ can be obtained using the probability of
neutrino detection given by Gaisser \& Grillo~(1987). The results of our
calculations, for the surface magnetic fields and initial periods of
pulsars specified by the models I, II, III, a the density factor $\xi = 1$,
are shown in Table~1 for the case of neutrinos arriving from directions close
to the horizon, i.e. not absorbed by the Earth (H), and for neutrinos which
arrive moving upward from the nadir direction and are partially absorbed (N)
(for absorption coefficients see Gandhi~2000). We investigated in more detail
the model III  for which the time dependent neutrino spectra are calculated
(see Figure~3).
The expected, time dependent, number of neutrinos detected by 1 km$^2$
detector in this case for  the horizontal and nadir directions are: 
$1.1\times 10^3$ (H) and 250 (N) (at $\Delta t = 10^4 - 10^5$ s), $2.7\times
10^3$ and 590 ($10^5 - 3\times 10^5$ s), $7.9\times 10^3$ and  $1.7\times
10^3$ ($3\times 10^5 - 10^6$ s), $1.2\times 10^4$ and  $2.5\times 10^3$ ($10^6
- 2\times 10^6$ s), $1.8\times 10^5$ and  $6.4\times 10^4$ ($2\times 10^6 -
10^7$ s), 400 and 150 ($10^7 - 3\times 10^7$ s). The highest detection rates
of neutrinos are expected  during about one-two months after supernova
explosion from the phase of particle interaction with the matter of the
envelope. 

If EHE CRs are produced by pulsars within our Galaxy, than the observed flux
of particles allows to constrain some free parameters of the considered model.
By comparing the observed flux of cosmic rays at $\sim 10^{20}$ eV with 
model estimations of the flux of iron nuclei Blasi et al.~(BEO) finds that
following condition should be fulfilled $\xi \epsilon Q/\tau_2
R_1^2B_{13}\approx 4\times 10^{-6}$, where $\xi$ is the efficiency for
accelerating particles defined similarly as in our paper, $\epsilon$ is the
fraction of pulsars which have the parameters required for particle
acceleration to $10^{20}$ eV, $Q$ is the trapping factor of particles within
the Galactic Halo, $\tau = 100\tau_2$ yr is the rate of neutron star
production, $R = 10R_1$ kpc is the  radius of the Galactic Halo, and $B_{13}
=0.1B_{12}$ is the pulsar surface magnetic field. For plausible parameters:
$\tau_2 = 1$, $R_1 = 3$ (required by the condition of isotropisation of EHE
CRs), $Q\sim 1$ (see recent calculations of the propagation of hadrons within
the Galaxy and Halo by Alvarez-Muniz, Engel \& Stanev~2001, Bednarek, Giller
\& Zieli\'nska~2001, and O'Neill, Olinto \& Blasi~2001), and the rate
of formation of neutron stars with required parameters equal to about 10\%
($\epsilon = 0.1$), we obtain the limit on the particle acceleration
efficiency $\xi\approx 10^{-4}$ for the model I, and  $\xi\approx 4\times
10^{-3}$ for the model III. These simple estimations combined with the results
of our calculations of the numbers of expected neutrinos, presented in Table~1,
show that some neutrinos might be observed in the 1 km$^2$ detector in the
case of the pulsar described by the model I and a few hundred of neutrinos in
the case of the model III. However, because of the steepness of the cosmic ray
spectrum such limits should be less restrictive for the pulsars
accelerating particles to lower energies  ($< 10^{20}$ eV). We conclude that
the detection of neutrinos from early phase of supernovae (or lack, thereof)
will put constraints on the recent models of extremely high energy cosmic ray
production in supernova explosion with formation of very energetic pulsar
(BEO; De Goubeia Dal Pino \& Lazarian~2000).

It is clear from  Table~1 that neutrinos from a  Crab-type pulsar located at
the distance of $\sim 2$ kpc (see our model II) might be observable by the 1
km$^2$ neutrino detector during the first year after pulsar formation if the
particle acceleration efficiency is $\xi > 3\times 10^{-3}$. Therefore it is
likely that a recent explosion of a supernova at a distance similar to that
of historical supernovae can also constrain the parameter $\xi$.

If the considered model works, then the whole population of pulsars created
in the Universe should contribute to the extragalactic neutrino background.
This could be detectable, because in such a case we do not need to be lucky to
find the pulsar within the Galaxy during such an early phase. This interesting
problem is considered in another paper (Bednarek~2001) in which we estimate
the extragalactic neutrino background from the population of pulsars with
parameters similar to those of classical radio pulsars formed in the Universe.

In fact, neutrinos can also be produced in later stages of supernova 
explosions, when the capturing of relativistic particles by the supernova
envelope is efficient. Such scenarios have been considered in the case of
proton acceleration in the pulsar's wind zone (Berezinsky \& Prilutsky~1978)
and in the case of hadron acceleration in the pulsar inner magnetosphere
(Bednarek \& Protheroe~2001, Protheroe, Bed\-na\-rek \& Luo~1998).  However,
such a production of neutrinos is less certain, since it is expected that
neutron stars can lose energy very efficiently via gravitational waves at
about one year after explosion due to the r-mode instabilities.
It has been argued that r-mode instabilities are not excited in the neutron
stars with the surface magnetic fields typical for magnetars, i.e.
$B >> 10^{13}$ G (Rezzolla, Lamb \& Shapiro~1999).  Detection of neutrinos
at these later times might militate against the r-mode instabilities as a
means of the production of gravitational radiation.  This in turn could have
implications for the likelihood of detection of gravitational radiation
associated with neutron star production.  On the other hand, a sharp cutoff
in the detected neutrino flux from a supernova could corroborate the
existence r-mode instabilities and lend credence to the possibility of
gravity wave generation and detection. Such a detection is of preeminent
theoretical interest.

We are grateful to an anonymous referee for comments and suggestions which
have improved the paper. WB thanks the the School for Computational Sciences
(SCS) at George Mason University at Fairfax (Virginia) for hospitality
during his visit. The research of WB was supported by the Polish KBN grant
No. 5P03D 025 21.

\clearpage
\begin{table} 
      \caption{Expected number of detected $\nu_\mu$} 
	 \begin{tabular}{|c|c|c|c|} 
	    \hline 
            \hline
	     & $P_{\rm ms} = 3$ & $P_{\rm ms} = 20$& $P_{\rm ms} = 10$\\ 
	     & $B_{12} = 4$ & $B_{12} = 4$ & $B_{12} = 100$\\ 
	     \hline 
	    N-$\gamma \rightarrow \nu_\mu$ (H)& $8.4\times 10^3$ & ~~~2.6 &
$2.4\times 10^4$\\   	    \hline 
	    Fe-M$\rightarrow \nu_\mu$ (H)      & $8.7\times 10^4$ & ~~11.3
 & $1.8\times 10^5$\\
 	    \hline
	    N-$\gamma \rightarrow \nu_\mu$ (N)& $2\times 10^3$  & ~~~1.1  &
$5.1\times 10^3$\\   	    \hline 
	    Fe-M$\rightarrow \nu_\mu$ (N)      & $3.4\times 10^4$  & ~~~8.6&
$6.4\times 10^4$\\
 	    \hline 
            \hline
	 \end{tabular} 
	 \label{tab1} 
     \end{table}

\end{document}